\documentclass[%
 reprint,
 amsmath,amssymb,
 aps,
floatfix,
]{revtex4-2}

\usepackage{graphicx}
\usepackage{dcolumn}
\usepackage{bm}

\usepackage{amsmath}
\usepackage{amsthm}
\usepackage{amssymb}
\usepackage{accents}
\usepackage{mathrsfs}
\usepackage{graphicx, subcaption}
\usepackage{float}
\DeclareMathOperator{\Tr}{Tr}

\DeclareMathOperator{\im}{Im}

\DeclareMathOperator{\Expv}{\mathbb{E}}

\newcommand{\wt}[1]{\tilde{#1}}

\DeclareFontFamily{OT1}{pzc}{}
\DeclareFontShape{OT1}{pzc}{m}{it}{ <-> s*[1.1] pzcmi7t }{}
\DeclareMathAlphabet{\mathpzc}{OT1}{pzc}{m}{it}

\makeatletter
\newcommand{\sbullet}{%
	\hbox{\fontfamily{lmr}\fontsize{.6\dimexpr(\f@size pt)}{0}\selectfont\textbullet}}

\makeatother

\numberwithin{theorem}{section} 
\usepackage{xcolor}[svgnames]

\newcommand{\M}{\mathcal{M}}
\newcommand{\oM}{\wt{\mathcal{M}}}

\newcommand{\m}{\mathfrak{m}}
\newcommand{\om}{\wt{\mathfrak{m}}}

\newcommand{\ms}{m}
\newcommand{\oms}{\wt{m}}

\newcommand{\s}{\mathfrak{s}}
\newcommand{\os}{\wt{\mathfrak{s}}}

\newcommand{\oq}{\wt{q}}

\begin{document}

\title{Eigenvector overlaps of sample covariance matrices with intersecting time periods}

\author{Volodymyr Riabov}\altaffiliation[Supported by ]{the ERC Advanced Grant "RMTBeyond"
No.~101020331.}
\affiliation{%
 Institute of Science and Technology Austria, Am Campus 1, 3400 Klosterneuburg, Austria\\
}%
\author{Konstantin Tikhonov}%
\author{Jean-Philippe Bouchaud}
 \altaffiliation[Also at ]{Académie des Sciences, Paris}
\affiliation{
Capital Fund Management, 23 rue de l'Université, 75007 Paris, France
}%

\date{\today}

\begin{abstract}
We compute exactly the overlap between the eigenvectors of two large empirical covariance matrices computed over intersecting time intervals, generalizing the results obtained previously for non-intersecting intervals. Our method relies on a particular form of Girko linearisation and extended local laws. We check our results numerically and apply them to financial data.  
\end{abstract}

\maketitle


\section{Introduction}

The computation of covariance matrices is the simplest tool to investigate the interactions that underlie the dynamics of complex systems, such as amorphous assemblies of molecules or hard spheres \cite{brito2010elementary}, neural networks (real \cite{almog2019uncovering} or synthetic \cite{louart2018random, atanasov2024scaling}), multi-species ecologies \cite{akjouj2024complex} or financial markets (for a review, see e.g. \cite{potters2020first} and refs therein). Often, the number of degrees of freedom $N$ is large (for example $N=500$ for the stocks belonging to the S\&P500). Now, the empirical determination of such large covariance matrices is difficult -- for at least two different reasons. One is that even in a stationary world, that is, a world described by an unknown underlying stochastic process with time independent parameters, empirical covariance matrices are soiled by a large amount of {\it measurement noise}, that only goes to zero as $\sqrt{q}$ with $q=N/T$ and $T$ is the amount of data points in the time direction. 

However, the assumption of a stationary world is, in many cases, unwarranted. In the case of financial markets, it is difficult to believe that data from 50 years ago can be used to reduce the parameter $q$ and lead to better estimates of the covariance matrix. The main difficulty is to disentangle measurement noise, inducing an apparent evolution of the empirical covariance matrix for two different periods over which it is computed, from any possible evolution of the ``true'' (or ``population'') $N \times N$ covariance matrix $C$. 

In Ref. \cite{bun2018overlaps}, the authors proposed to use the overlap between the eigenvectors of the empirical covariance matrices computed over two different periods to distinguish between the pure noise hypothesis and the genuine evolution hypothesis, a subject first discussed in \cite{allez2012eigenvector}. More precisely, assuming that the population covariance $C$ is time {\it independent}, one can exactly compute the mean overlap $\mathbb{E}[|\langle {\bf u}_j, \widetilde {\bf u}_\ell \rangle|^2]$ of eigenvectors ${\bf u}_j$, $\widetilde {\bf u}_\ell$ corresponding to the two empirical covariances matrices, associated to eigenvalues $\lambda_j$ and $\tilde \lambda_\ell$. Such a prediction then provides a null-hypothesis benchmark that allows one to test for non-stationarity (see also \cite{bouchaud2022} for an alternative proposal). It also provides an intuitive interpretation the Ledoit-Péché formula for eigenvalue cleaning \cite{ledoit2011eigenvectors, bun2018overlaps, potters2020first}.

However, the Random Matrix Theory (RMT) approach of Ref. \cite{bun2018overlaps} relied on the fact that the two considered time periods are non-intersecting. However, it is often useful to consider cases where time periods do overlap, for example when one uses sliding time windows. This could potentially help identify sudden regime shifts. In this (rather technical) note, we extend the result of \cite{bun2018overlaps} to arbitrary overlapping periods. We also test our new formulas on both synthetic and real financial data.

\section{Model}

Our goal is to compute the averaged eigenvector overlap between two manipulatively deformed $N\times N$ Wishart matrices $W$ and $\wt{W}$ that are generated from a partially overlapping datasets of the form 
\begin{equation}
    \begin{split}
        W &:= q \, C^{1/2} \bigl(XX^\top + BB^\top\bigr)C^{1/2}\,\\ \wt{W} &:= \oq \, C^{1/2} \bigl(\wt{X}\wt{X}^\top + BB^\top\bigr)C^{1/2},
    \end{split}
\end{equation}
where $X$, $\wt X$ and $B$ are $N \times T$, $N \times \wt{T}$, and $N \times T_B$, respectively, random matrices with independent centered Gaussian entries of variance $N^{-1}$. We consider $X, \wt X$ and $B$ to be jointly independent. Here
\begin{equation}
	q := \frac{N}{T + T_B}, \qquad \oq  := \frac{N}{\wt{T} + T_B},
\end{equation} 
where $T_B$ is the length of the overlapping region, and $T+T_B$, $\wt{T}+T_B$ the total length of the two samples. 

By spectral decomposition, the eigenvector overlaps (up to $z, \wt{z}$-dependent normalization) can be directly extracted from the trace of the product of two resolvents (see Eq. (5) in \cite{bun2018overlaps}),
\begin{equation} \label{eq:overlap}
	\psi(z,\wt{z}) := \tau \left[ \bigl( W - z I_N \bigr)^{-1} \bigl( \wt{W} - \wt{z} I_N \bigr)^{-1} \right],
\end{equation}
where $\tau[\, \cdot \,] := N^{-1} \Tr [\, \cdot \,]$, and $z, \wt{z} \in \mathbb{C}\backslash \mathbb{R}$ are spectral parameters. Here $I_d$ denotes the $d\times d$ identity matrix, but in the sequel, we often identify scalars with corresponding multiples of the identity matrix where no confusion may arise.

Since $W$ and $\wt{W}$ share the component $W_B$, and hence are correlated, the customary single-resolvent local laws do not directly apply. Therefore, the two Wishart matrices have to be combined into a large block random matrix, and the quantity \eqref{eq:overlap} can then be studied using two-resolvent local laws. 

\section{Linearization and Extended Local Laws}

Here we give the main idea of our calculation and relegate heavy details to the Supplementary Material (SM) section. We first construct the combined Girko linearization $H$ of $W$ and $\wt{W}$ in the form
\begin{equation} \label{eq:H_def}
	H := \begin{pmatrix}
		0 & Y\\
		Y^\top & 0
	\end{pmatrix},
\end{equation}
where $Y$ is an $(2N) \times (T + \wt{T} + 2T_B)$ matrix of the form
\begin{equation*}
	Y := \begin{pmatrix}
		\sqrt{q}\, C^{1/2}X & 0 & \sqrt{q}\,C^{1/2}B & 0 \\
		0 & \sqrt{\oq }\,C^{1/2}\wt{X} & 0 & \sqrt{\oq }\, C^{1/2}B
	\end{pmatrix}.
\end{equation*}
In particular, $H$ is a symmetric $L \times L$ matrix with
\begin{equation} \label{eq:L_def}
	L := (2 N + T + \wt T + 2 T_B).
\end{equation} 
Observe that $HH^\top = H^2$ satisfies
\begin{equation} \label{eq:H^2}
	HH^\top = \begin{pmatrix}
		\mathcal{H} & 0\\
		0 &  \star
	\end{pmatrix}, 
	\quad 
	\mathcal{H} := \begin{pmatrix}
		W & 0 \\
		0 & \wt{W}
	\end{pmatrix}~,
\end{equation}
where here and in the following `$\star$' indicates blocks that are irrelevant for our purpose. 

Hence, with $G(z) := (H-z)^{-1}$ and $\mathcal{G}(z) := (\mathcal{H} - z)^{-1}$ denoting the resolvents of $H$ and $\mathcal{H}$, it is straightforward to check (e.g., using spectral decomposition) that $G(z)$ satisfies
\begin{equation}
	G(z) = \begin{pmatrix}
		z\mathcal{G}(z^2) & \star\\
		\star & \star
	\end{pmatrix}.
\end{equation}
Therefore, the quantity \eqref{eq:overlap} can be expressed as
\begin{equation} \label{eq:full_overlap}
        \psi(z,\wt{z}) = \frac{1}{w \wt{w}} \, {\tau \Bigl[  G(w) P   G(\wt{w}) P^\top \Bigr]},
\end{equation}
with 
\begin{equation}
	P := \begin{pmatrix}
		\mathcal{P} & 0 \\
		0 & 0
	\end{pmatrix}, \quad \mathcal{P} := \begin{pmatrix}
	0 & I_N\\
	0 & 0
	\end{pmatrix},
\end{equation}
and where $w_j := \sqrt{z_j}$ for $j=1,2$. We choose the branch of the square-root function cut along $[-\infty, 0)$. We conclude that it suffices to estimate $\tau \bigl[  G(w) P   G(\wt{w}) P^\top \bigr]$.

The resolvent $G(w)$ of the large random matrix exhibits a strong concentration of measure, which means that its behavior is approximately deterministic, down to the smallest spectral scales $|\im w| \gg N^{-1}$. Provided $\Expv [H] = 0$, the deterministic approximation to a single resolvent $G(w)$ at a spectral parameter $w$ is given by a matrix $M(w)$ that solves the Dyson equation \cite{Alt2020energy}
\begin{equation} \label{eq:MDE}
	- \bigl(M(w)\bigr)^{-1} = w + S \bigl[M(w)\bigr], \quad \bigl( \im M(w) \bigr) \bigl( \im w\bigr) > 0,
\end{equation}
where the positivity condition is understood in the sense of quadratic forms, and the imaginary part of matrix $M$ is given by $\tfrac{1}{2i}(M - M^\star)$, with $(\cdot)^\star$ denoting the Hermitian conjugate. 
Here $S : \mathbb{C}^{L\times L} \to \mathbb{C}^{L\times L}$ denotes the (super-) operator, called the \textit{self-energy} of $H$, defined through its action on $L\times L$ deterministic matrices $A$ by
\begin{equation} \label{eq:self-energy}
	S[A] := \Expv \bigl[H\, A\, H\bigr].
\end{equation}
The solution to the matrix Dyson equation \eqref{eq:MDE} is unique and corresponds to the Stieltjes transform of a compactly supported matrix-valued measure on the real line under minor assumptions of $S$.


A recent further refinement of the local law framework is the introduction of multi-resolvent local laws \cite{Cipolloni2022Optimal}, which establish a similar concentration of measure results for products of multiple resolvents interlaced with deterministic observables (\textit{resolvent chains}). In particular, a product of two resolvents is well approximated \footnote{
To the best of our knowledge, the two-resolvent local law \eqref{eq:2Glaw} for sample covaraince matrices was first established in \cite{lin2024eigenvector}: their result is suboptimal and restricted to non-overlapping samples, and the error is suboptimal. However, with our dynamical proof techniques \cite{erdHos2024eigenstate}, the optimal result is attainable even in the correlated case \cite{erdHos2024cusp}. } by
\begin{equation} \label{eq:2Glaw}
	G(w) A G(\wt{w})  \approx   M(w, A, \wt{w}),
\end{equation}
where $M(w, A, \wt{w})$ is an explicit deterministic matrix, which admits the expression \cite{Cipolloni2022overlap}
\begin{equation} \label{eq:M12_def}
	M(z, A, \wt{z}) := \mathcal{B}_{w, \wt{w}}^{-1} \bigl[M(w) A M(\wt{w})\bigr],
\end{equation}
where $\mathcal{B}_{w, \wt{w}} [A] := A - M(w) S [A] M(\wt{w})$ is the two-body \textit{stability operator} associated with the equation \eqref{eq:MDE}. In view of
(\ref{eq:full_overlap}) suggests, we have to compute $M(w, P, \wt{w})$, which we denote by $M^P$ for brevity. Equation \eqref{eq:M12_def} is equivalent to 
\begin{equation}
\label{eq:MP_eq}
	M^P - M(w) S\bigl[ M^P \bigr] M(\wt{w}) :=  M(w) P M(\wt{w}),
\end{equation}
where $M(w)$ is the solution to Eq.~\eqref{eq:MDE}. Note that, in general, $M(w, A, \wt{w}) \neq M(w) A M(\wt{w})$, since the two-body $M$-term \eqref{eq:M12_def} has to correctly capture the correlations between the two resolvents.

\section{Computing the Deterministic Overlap}
In this section, we compute $\psi_{\rm d}(z, \wt{z})$, defined as 
\begin{equation} \label{eq:M_overlap}
	\psi_{\rm d}(z, \wt{z}) :=  \frac{1}{w \wt{w}}\tau \bigl[ M(w, P, \wt{w}) P^\top \bigr],
\end{equation}
with $w := \sqrt{z}$, $\wt{w} := \sqrt{\wt{z}}$, which is the deterministic counterpart to \eqref{eq:full_overlap} by \eqref{eq:2Glaw}, in the sense that 
\begin{equation}
	\psi(z, \wt{z}) \approx \psi_{\rm d}(z, \wt{z}).
\end{equation}

	
	

We start with solving the Dyson equation \eqref{eq:MDE}.
By using expressions \eqref{eq:S}--\eqref{eq:T1R1_def} from the SM, we can show that $M(w)$ is a block-diagonal matrix with $\mathcal{M}(w)$, $\wt{\mathcal{M}}(w)$, $\m(w) I_T$, $\om(w) I_{\wt{T}}$, $\m(w) I_{T_B}$, $\om(w) I_{T_B}$ on the main diagonal and
\begin{equation} \label{eq:Ms}
    \begin{split}
        &\left\{ 
	\begin{array}{l}
		\smallskip
		-\mathcal{M}(w)^{-1} = w +  \m(w) C \\
		\smallskip
		-\m(w)^{-1}= w +  q \,\tau \bigl[\mathcal{M}(w)C \bigr]\\
	\end{array} 
	\right.~,\\
	&\left\{ 
	\begin{array}{l}
		\smallskip
		-\wt{\mathcal{M}}(w)^{-1} = z + \om(w)  C \\
		\smallskip
		-\om(w)^{-1}= w +  \oq \,\tau \bigl[\wt{\mathcal{M}}(w) C\bigr]
	\end{array} 
	\right. ~.
    \end{split}
\end{equation}
It follows easily that the functions $\M(w)$, $\m(w)$ satisfy
\begin{align} \label{eq:gammas}
    \m(w) &= \frac{q - 1}{w} + q \tau\bigl[ \M(w) \bigr],\nonumber \\
    \mathcal{M}(w) &= -\frac{ 1 }{\m(w)C + w},
\end{align}
and the relation for $\oM$, $\om$ is analogous.
These equations correspond to the celebrated Marc\v{e}nko--Pastur law and its extension to general population covariance matrices \cite{ledoit2011eigenvectors, marvcenko1967distribution}. Following Ref.~\cite{bun2018overlaps}, it is convenient to introduce the functions $\ms \equiv \ms(z)$ and $\oms \equiv \oms(\wt{z})$ of the spectral parameters $z$ and $\wt{z}$, defined as
\begin{equation} \label{eq:ms}
    \ms :=- \frac{\m}{w}, \quad \oms:= -\frac{\om}{\wt{w}}, \quad w := \sqrt{z},\, \wt{w} := \sqrt{\wt{z}}.
\end{equation}
It follows from \eqref{eq:gammas} and \eqref{eq:ms} that
\begin{equation}
	\tau \bigl[ \M \bigr]  = w\frac{1 - q   - z \ms}{q z}, \quad \tau \bigl[ \oM \bigr]  = \wt{w}\frac{1 - \oq - \wt{z} \oms}{ \oq \wt{z}}.
\end{equation}

Having the Dyson equation solved, we may proceed with the main Eq.~\eqref{eq:MP_eq}. As we show in the SM, the object $M^P$ can be written in the block-diagonal form and this equation reduces to
\begin{equation} \label{eq:F12_expr}
    \begin{split}
	\mathcal{F}_{12} &- t \sqrt{q\oq} \, \tau \bigl[ \mathcal{F}_{12}C \bigr] \m \om\, \mathcal{M} C \wt{\mathcal{M}} = \mathcal{M} \wt{\mathcal{M}}, 
    \end{split}
\end{equation}
where $\mathcal{F}_{12}$ is a certain $N\times N$ sub-matrix of $M^P$ and we abbreviate $\M := \mathcal{M}(w)$, $\m := \m(w)$, and $\oM := \oM(\wt{w})$, $\om := \om(\wt{w})$.  The parameter $t := N^{-1} T_B\sqrt{q\oq} \in [0,1]$  in \eqref{eq:F12_expr} regulates the strength of the coupling between $W$ and $\wt{W}$, that is
\begin{equation}
	t = \mathrm{Corr}\bigl(XX^\top + BB^\top, \wt{X}\wt{X}^\top + BB^\top\bigr)
\end{equation}
and correlation is computed entry-wise.

The same sub-matrix $\mathcal{F}_{12}$ is sufficient to compute the overlap function, since as we show in the SM:
\begin{equation}
	\tau \bigl[ M^P P^\top\bigr] = \tau \bigl[ \mathcal{F}_{12}\bigr].
\end{equation}
Multiplying both sides of \eqref{eq:F12_expr} by $C$ and taking the normalized trace, we obtain
\begin{equation}
	\tau \bigl[ \mathcal{F}_{12}C \bigr] = \frac{\tau \bigl[ \mathcal{M} \wt{\mathcal{M}}C \bigr]}{1 - t \sqrt{q\oq} \, \m \om\,\tau \bigl[  \mathcal{M} C \wt{\mathcal{M}}C \bigr]}.
\end{equation}
The fact that the denominator is not singular can be established by analyzing the stability operator $\mathcal{B}_{w, \wt{w}}$, defined in \eqref{eq:M12_def}.
Therefore, the trace of $\mathcal{F}_{12}$ is given by
\begin{equation}
	\tau \bigl[ \mathcal{F}_{12} \bigr] = \tau \bigl[ \mathcal{M} \wt{\mathcal{M}} \bigr] +  \frac{t\sqrt{q\oq} \, \m \om\, \left(\tau \bigl[ \mathcal{M} C \wt{\mathcal{M}} \bigr]\right) ^2}{1 - t \sqrt{q\oq}\, \m \om\,\tau \bigl[ \mathcal{M} C \wt{\mathcal{M}}C \bigr]}.
\end{equation}
Here we used that $\mathcal{M}$ and $\wt{\mathcal{M}}$ are resolvents of $C$, and hence commute with it.


Furthermore, the resolvent identities imply that 
\begin{equation}
	\tau \bigl[ \M C \oM \bigr] = \frac{1}{w \wt{w}}\frac{1}{\ms - \oms}\biggl(\frac{1 - z\ms}{q} - \frac{1 - \wt{z}\oms}{\oq}\biggr),
\end{equation}
\begin{equation}
	\tau \bigl[ \M C \oM C \bigr] = \frac{1}{w \wt{w}}\frac{1}{\ms - \oms}\biggl(\frac{1 - z\ms}{q \ms} - \frac{1 - \wt{z}\oms}{\oq \oms} \biggr).
\end{equation}

These expressions allow us to derive our central result for the averaged trace of resolvent product $\psi_{\rm d}(z, \wt{z})$, defined in \eqref{eq:M_overlap} and which in turn give access to the eigenvector overlaps \cite{bun2018overlaps}: 
\begin{equation} \label{eq:psi}
\boxed{
    \begin{split}
	\psi_{\rm d}(z, \wt{z}) =&~ \frac{1}{z \wt{z}}\Biggl[1 + \frac{\ms\oms}{\ms - \oms}\biggl(\frac{1 - \wt{z}\oms}{\oq\ms}-\frac{1 - z\ms}{q\oms}\biggr) \\
        &+ \frac{t\sqrt{q\oq} \,  \frac{\ms\oms}{(\ms - \oms)^2}\, \bigl(\frac{1 - z\ms}{q} - \frac{1 - \wt{z}\oms}{\oq}\bigr) ^2}
	{1 - t\sqrt{q\oq} \frac{\ms \oms}{\ms - \oms}\bigl(\frac{1 - z\ms}{q \ms} - \frac{1 - \wt{z}\oms}{\oq \oms}  \bigr) }\Biggr].
    \end{split}
    }
\end{equation}
Note that for $t=0$,  $\psi_{\rm d}(z, \wt{z})$ boils down to the result of \cite{bun2018overlaps} for the case of non-intersecting time periods.

\section{Numerics \& Conclusion}

\begin{figure}[t]
	{\centering	\includegraphics[width=.49\textwidth]{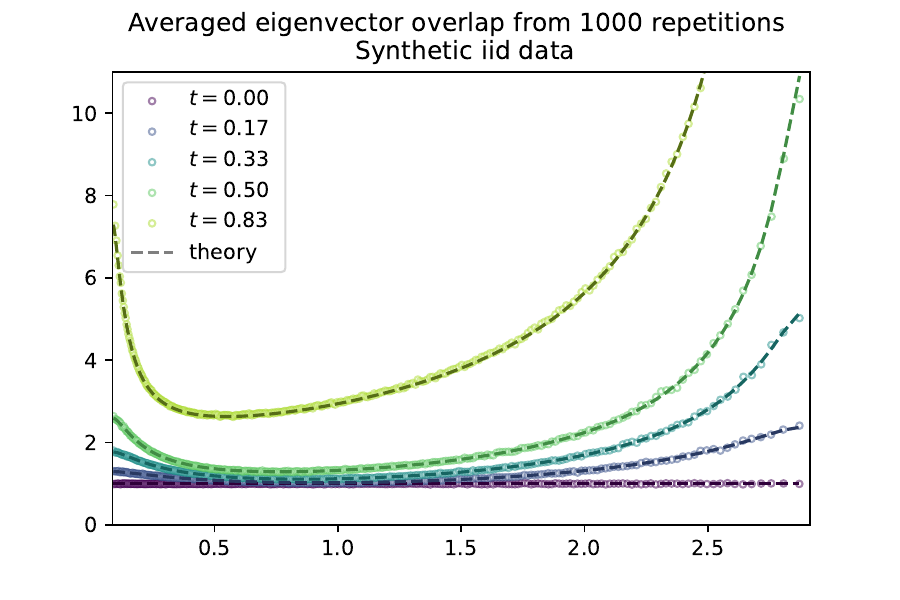}}
	\caption{\small{Depicted are averaged eigenvector overlaps $\Phi(\lambda)$ of covariance matrices $W$ and $\wt{W}$ as a function of $\lambda$. The population covariance is $C = I_N$, and we collected at least $1000$ realizations for every value of $t$. The theoretical curves computed using \eqref{eq:sp} at corresponding values of $t$ are shown as dashed lines. } }
	\label{fig:iid}
\end{figure}

In this section, we present numerical experiments that illustrating our theoretical results for the eigenvector overlaps $\Phi(\lambda,\wt{\lambda}) = N\mathbb{E}[|\langle {\bf u}_\lambda, \widetilde {\bf u}_{\wt{\lambda}} \rangle|^2]$. This quantity can be deduced from $\psi_{\rm d}(z, \wt{z})$ in \eqref{eq:psi} via Sokhotski-Plemelj identity,
\begin{equation} \label{eq:sp}
\Phi(\lambda, \wt{\lambda})=\frac{1}{2\pi^2\rho(\lambda)\rho(\wt{\lambda})}\mathrm{Re}\left[\psi_{\rm d}(z, \wt{z}^*)-\psi_{\rm d}(z, \wt{z})\right],
\end{equation}
where $z=\lambda+i\eta$, $\wt{z}=\wt{\lambda}+i\wt{\eta}$ with $\eta, \wt{\eta} \to +0$, and $\rho$ is the spectral density of $C$.

For simplicity, in this numerical study we only consider self-overlap $\Phi(\lambda)$,
\begin{equation} \label{eq:overlap_num}
    \begin{split}
       \Phi(\lambda_j) &:= \frac{1}{Z_j} \sum_{\ell = 1}^N \frac{\bigl\lvert\bigl\langle \bm u_j, \wt{\bm u}_\ell \bigr\rangle\bigr\rvert^2 }{\bigl( \lambda_j - \wt{\lambda}_\ell \bigr)^2 + \eta^2},
    \\ Z_j &:= \frac1N \sum_{\ell = 1}^N \frac{1}{\bigl( \lambda_j - \wt{\lambda}_\ell \bigr)^2 + \eta^2},  \quad \eta := \frac{1}{\sqrt{N}}.
    \end{split}
\end{equation}
This definition corresponds to the right-hand side of \eqref{eq:sp} with $\eta = N^{-1/2}$ and $\wt{\eta} = N^{-1}$ when comparing theoretical results to numerical ones.

For numerical studies, we consider the covariance matrices of the size $N=300$ and width parameters $q = \oq = 1/2$. First, we focus on synthetic data with independent identically distributed (iid) entries in Figure~\ref{fig:iid}, averaging over $1000$ realizations of randomness for each value of $t$. The agreement with our theory is excellent for all $t$, even for values of $N=300$ that are not extremely large.

\begin{figure}[t]
		\includegraphics[width=.49\textwidth]{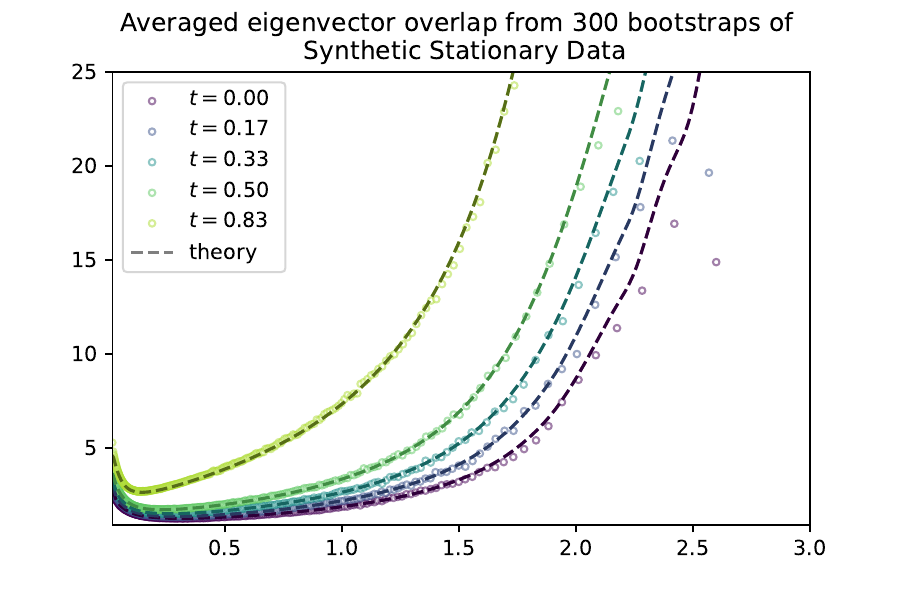}
	\caption{\small{Averaged eigenvector overlaps $\Phi(\lambda)$ of covariance matrices $W$ and $\wt{W}$ computed from a time-series of Gaussian returns with population correlation $C = C_0$, where $C_0$ is the empirical correlation matrix of the stock returns. The dashed lines are the theoretical curves computed using \eqref{eq:sp}.}} 
	\label{fig:station}
\end{figure}

As a next step, we compute the overlaps based on real financial time-series data. In particular, we consider a time series of returns of $N=300$ US stocks in the period of $2004-2013$. To begin with, we compute the full-sample correlation matrix $C_0$, sample the (fictitious)  Gaussian returns from this correlation matrix, thus removing the possible effects of non-stationarity altogether. The results are shown in Figure~\ref{fig:station}. The agreement between the actual overlaps and those predicted by \eqref{eq:sp} is still very good, except at the larger eigenvalues: this discrepancy appears from the fact that consecutive eigenvalue gaps are large in this energy range for the empirical correlation matrix $C_0$ and the two-resolvent local law error is non-negligible (this error is $\propto \rho^{-1}(z)$, where $\rho(z)$ is the density of states regularized on the scale $|\im z|=\eta$).

\begin{figure}[t]
\centering
		\includegraphics[width=.49\textwidth]{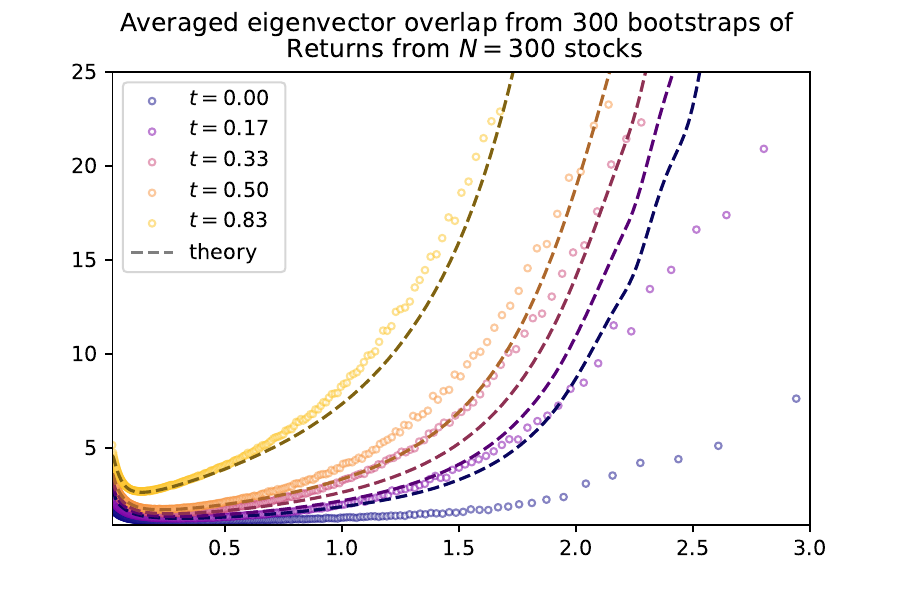}
	\caption{\small{Averaged eigenvector overlaps $\Phi(\lambda)$ of covariance matrices $W$ and $\wt{W}$ obtained from the time series of the $N=300$ stocks using the bootstrapping procedure. The dashed lines are the theoretical curves computed using \eqref{eq:sp}. Note the strong discrepancies that appear when $t$ decreases, indicative of non-stationarity.}  }
	\label{fig:real}
\end{figure}

Finally, we compute the overlaps directly on the real returns time series. For the results in Figure~\ref{fig:real}, we apply a bootstrapping procedure by randomly sub-sampling returns to generate $W$ and $\wt{W}$ with $T=600$ as in \cite[Fig. 3]{bun2018overlaps}.
Observe that the agreement between the scattered points with the theoretical curves in Fig.~\ref{fig:real} deteriorates compared to Fig.~\ref{fig:station}. For small values of $t$, the displacement in Figure~\ref{fig:real} is due to the strong non-stationarity of the underlying population covariance on a timescale comparable to $N/q = 600$. Note that this effect diminishes as $t$ grows, since the samples generating $W$ and $\wt{W}$ become increasingly influenced by the shared block $B$. None of these effects are present for stationary samples as illustrated in Fig.~\ref{fig:station}.

Note that the central result Eq. \eqref{eq:psi} is expected to hold, for large $N$, for non-Gaussian data provided that their second moment exists (generalization to random entries with finite moments is straightforward). However, finite-size effects may also mimic non-stationarity and are expected to be more pronounced in the presence of heavy tails, as is often the case for financial data, see e.g., \cite{biroli2007student, potters2020first}. A more detailed investigation of this issue would be of considerable interest; in particular, heavy-tail effects can be mitigated by employing alternative correlation measures, such as Spearman or Kendall coefficients \cite{espana2024kendall}. Determining whether our results extend to such settings remains an open problem. It would also be of great interest to generalize the alternative approach proposed in \cite{bouchaud2022} to the case of overlapping time periods, which appears to be tractable using different RMT methods. 

Our proof of Eq. \eqref{eq:psi} relies on the robust but highly technical framework of multi-resolvent local laws and underlying Dyson equations. The authors suspect that a more direct derivation exists in the Gaussian case, but so far it has not been found. 

\bibliography{apssamp}

\begin{thebibliography}{20}%
\makeatletter
\providecommand \@ifxundefined [1]{%
 \@ifx{#1\undefined}
}%
\providecommand \@ifnum [1]{%
 \ifnum #1\expandafter \@firstoftwo
 \else \expandafter \@secondoftwo
 \fi
}%
\providecommand \@ifx [1]{%
 \ifx #1\expandafter \@firstoftwo
 \else \expandafter \@secondoftwo
 \fi
}%
\providecommand \natexlab [1]{#1}%
\providecommand \enquote  [1]{``#1''}%
\providecommand \bibnamefont  [1]{#1}%
\providecommand \bibfnamefont [1]{#1}%
\providecommand \citenamefont [1]{#1}%
\providecommand \href@noop [0]{\@secondoftwo}%
\providecommand \href [0]{\begingroup \@sanitize@url \@href}%
\providecommand \@href[1]{\@@startlink{#1}\@@href}%
\providecommand \@@href[1]{\endgroup#1\@@endlink}%
\providecommand \@sanitize@url [0]{\catcode `\\12\catcode `\$12\catcode `\&12\catcode `\#12\catcode `\^12\catcode `\_12\catcode `\%12\relax}%
\providecommand \@@startlink[1]{}%
\providecommand \@@endlink[0]{}%
\providecommand \url  [0]{\begingroup\@sanitize@url \@url }%
\providecommand \@url [1]{\endgroup\@href {#1}{\urlprefix }}%
\providecommand \urlprefix  [0]{URL }%
\providecommand \Eprint [0]{\href }%
\providecommand \doibase [0]{https://doi.org/}%
\providecommand \selectlanguage [0]{\@gobble}%
\providecommand \bibinfo  [0]{\@secondoftwo}%
\providecommand \bibfield  [0]{\@secondoftwo}%
\providecommand \translation [1]{[#1]}%
\providecommand \BibitemOpen [0]{}%
\providecommand \bibitemStop [0]{}%
\providecommand \bibitemNoStop [0]{.\EOS\space}%
\providecommand \EOS [0]{\spacefactor3000\relax}%
\providecommand \BibitemShut  [1]{\csname bibitem#1\endcsname}%
\let\auto@bib@innerbib\@empty
\bibitem [{\citenamefont {Brito}\ \emph {et~al.}(2010)\citenamefont {Brito}, \citenamefont {Dauchot}, \citenamefont {Biroli},\ and\ \citenamefont {Bouchaud}}]{brito2010elementary}%
  \BibitemOpen
  \bibfield  {author} {\bibinfo {author} {\bibfnamefont {C.}~\bibnamefont {Brito}}, \bibinfo {author} {\bibfnamefont {O.}~\bibnamefont {Dauchot}}, \bibinfo {author} {\bibfnamefont {G.}~\bibnamefont {Biroli}},\ and\ \bibinfo {author} {\bibfnamefont {J.-P.}\ \bibnamefont {Bouchaud}},\ }\bibfield  {title} {\bibinfo {title} {Elementary excitation modes in a granular glass above jamming},\ }\href@noop {} {\bibfield  {journal} {\bibinfo  {journal} {Soft Matter}\ }\textbf {\bibinfo {volume} {6}},\ \bibinfo {pages} {3013} (\bibinfo {year} {2010})}\BibitemShut {NoStop}%
\bibitem [{\citenamefont {Almog}\ \emph {et~al.}(2019)\citenamefont {Almog}, \citenamefont {Buijink}, \citenamefont {Roethler}, \citenamefont {Michel}, \citenamefont {Meijer}, \citenamefont {Rohling},\ and\ \citenamefont {Garlaschelli}}]{almog2019uncovering}%
  \BibitemOpen
  \bibfield  {author} {\bibinfo {author} {\bibfnamefont {A.}~\bibnamefont {Almog}}, \bibinfo {author} {\bibfnamefont {M.~R.}\ \bibnamefont {Buijink}}, \bibinfo {author} {\bibfnamefont {O.}~\bibnamefont {Roethler}}, \bibinfo {author} {\bibfnamefont {S.}~\bibnamefont {Michel}}, \bibinfo {author} {\bibfnamefont {J.~H.}\ \bibnamefont {Meijer}}, \bibinfo {author} {\bibfnamefont {J.~H.}\ \bibnamefont {Rohling}},\ and\ \bibinfo {author} {\bibfnamefont {D.}~\bibnamefont {Garlaschelli}},\ }\bibfield  {title} {\bibinfo {title} {Uncovering functional signature in neural systems via random matrix theory},\ }\href@noop {} {\bibfield  {journal} {\bibinfo  {journal} {PLoS computational biology}\ }\textbf {\bibinfo {volume} {15}},\ \bibinfo {pages} {e1006934} (\bibinfo {year} {2019})}\BibitemShut {NoStop}%
\bibitem [{\citenamefont {Louart}\ \emph {et~al.}(2018)\citenamefont {Louart}, \citenamefont {Liao},\ and\ \citenamefont {Couillet}}]{louart2018random}%
  \BibitemOpen
  \bibfield  {author} {\bibinfo {author} {\bibfnamefont {C.}~\bibnamefont {Louart}}, \bibinfo {author} {\bibfnamefont {Z.}~\bibnamefont {Liao}},\ and\ \bibinfo {author} {\bibfnamefont {R.}~\bibnamefont {Couillet}},\ }\bibfield  {title} {\bibinfo {title} {A random matrix approach to neural networks},\ }\href@noop {} {\bibfield  {journal} {\bibinfo  {journal} {The Annals of Applied Probability}\ }\textbf {\bibinfo {volume} {28}},\ \bibinfo {pages} {1190} (\bibinfo {year} {2018})}\BibitemShut {NoStop}%
\bibitem [{\citenamefont {Atanasov}\ \emph {et~al.}(2024)\citenamefont {Atanasov}, \citenamefont {Zavatone-Veth},\ and\ \citenamefont {Pehlevan}}]{atanasov2024scaling}%
  \BibitemOpen
  \bibfield  {author} {\bibinfo {author} {\bibfnamefont {A.}~\bibnamefont {Atanasov}}, \bibinfo {author} {\bibfnamefont {J.~A.}\ \bibnamefont {Zavatone-Veth}},\ and\ \bibinfo {author} {\bibfnamefont {C.}~\bibnamefont {Pehlevan}},\ }\bibfield  {title} {\bibinfo {title} {Scaling and renormalization in high-dimensional regression},\ }\href@noop {} {\bibfield  {journal} {\bibinfo  {journal} {arXiv preprint arXiv:2405.00592}\ } (\bibinfo {year} {2024})}\BibitemShut {NoStop}%
\bibitem [{\citenamefont {Akjouj}\ \emph {et~al.}(2024)\citenamefont {Akjouj}, \citenamefont {Barbier}, \citenamefont {Clenet}, \citenamefont {Hachem}, \citenamefont {Ma{\"\i}da}, \citenamefont {Massol}, \citenamefont {Najim},\ and\ \citenamefont {Tran}}]{akjouj2024complex}%
  \BibitemOpen
  \bibfield  {author} {\bibinfo {author} {\bibfnamefont {I.}~\bibnamefont {Akjouj}}, \bibinfo {author} {\bibfnamefont {M.}~\bibnamefont {Barbier}}, \bibinfo {author} {\bibfnamefont {M.}~\bibnamefont {Clenet}}, \bibinfo {author} {\bibfnamefont {W.}~\bibnamefont {Hachem}}, \bibinfo {author} {\bibfnamefont {M.}~\bibnamefont {Ma{\"\i}da}}, \bibinfo {author} {\bibfnamefont {F.}~\bibnamefont {Massol}}, \bibinfo {author} {\bibfnamefont {J.}~\bibnamefont {Najim}},\ and\ \bibinfo {author} {\bibfnamefont {V.~C.}\ \bibnamefont {Tran}},\ }\bibfield  {title} {\bibinfo {title} {Complex systems in ecology: a guided tour with large lotka--volterra models and random matrices},\ }\href@noop {} {\bibfield  {journal} {\bibinfo  {journal} {Proceedings of the Royal Society A}\ }\textbf {\bibinfo {volume} {480}},\ \bibinfo {pages} {20230284} (\bibinfo {year} {2024})}\BibitemShut {NoStop}%
\bibitem [{\citenamefont {Potters}\ and\ \citenamefont {Bouchaud}(2020)}]{potters2020first}%
  \BibitemOpen
  \bibfield  {author} {\bibinfo {author} {\bibfnamefont {M.}~\bibnamefont {Potters}}\ and\ \bibinfo {author} {\bibfnamefont {J.-P.}\ \bibnamefont {Bouchaud}},\ }\href@noop {} {\emph {\bibinfo {title} {A first course in random matrix theory: for physicists, engineers and data scientists}}}\ (\bibinfo  {publisher} {Cambridge University Press},\ \bibinfo {year} {2020})\BibitemShut {NoStop}%
\bibitem [{\citenamefont {Bun}\ \emph {et~al.}(2018)\citenamefont {Bun}, \citenamefont {Bouchaud},\ and\ \citenamefont {Potters}}]{bun2018overlaps}%
  \BibitemOpen
  \bibfield  {author} {\bibinfo {author} {\bibfnamefont {J.}~\bibnamefont {Bun}}, \bibinfo {author} {\bibfnamefont {J.-P.}\ \bibnamefont {Bouchaud}},\ and\ \bibinfo {author} {\bibfnamefont {M.}~\bibnamefont {Potters}},\ }\bibfield  {title} {\bibinfo {title} {Overlaps between eigenvectors of correlated random matrices},\ }\href {https://doi.org/10.1103/PhysRevE.98.052145} {\bibfield  {journal} {\bibinfo  {journal} {Physical Review E}\ }\textbf {\bibinfo {volume} {98}},\ \bibinfo {pages} {052145} (\bibinfo {year} {2018})}\BibitemShut {NoStop}%
\bibitem [{\citenamefont {Allez}\ and\ \citenamefont {Bouchaud}(2012)}]{allez2012eigenvector}%
  \BibitemOpen
  \bibfield  {author} {\bibinfo {author} {\bibfnamefont {R.}~\bibnamefont {Allez}}\ and\ \bibinfo {author} {\bibfnamefont {J.-P.}\ \bibnamefont {Bouchaud}},\ }\bibfield  {title} {\bibinfo {title} {Eigenvector dynamics: general theory and some applications},\ }\href@noop {} {\bibfield  {journal} {\bibinfo  {journal} {Physical Review E—Statistical, Nonlinear, and Soft Matter Physics}\ }\textbf {\bibinfo {volume} {86}},\ \bibinfo {pages} {046202} (\bibinfo {year} {2012})}\BibitemShut {NoStop}%
\bibitem [{\citenamefont {Bouchaud}\ \emph {et~al.}(2022)\citenamefont {Bouchaud}, \citenamefont {Mastromatteo}, \citenamefont {Potters},\ and\ \citenamefont {Tikhonov}}]{bouchaud2022}%
  \BibitemOpen
  \bibfield  {author} {\bibinfo {author} {\bibfnamefont {J.-P.}\ \bibnamefont {Bouchaud}}, \bibinfo {author} {\bibfnamefont {I.}~\bibnamefont {Mastromatteo}}, \bibinfo {author} {\bibfnamefont {M.}~\bibnamefont {Potters}},\ and\ \bibinfo {author} {\bibfnamefont {K.}~\bibnamefont {Tikhonov}},\ }\bibfield  {title} {\bibinfo {title} {Excess out-of-sample risk and fleeting modes},\ }\href {https://doi.org/10.54946/wilm.11056} {\bibfield  {journal} {\bibinfo  {journal} {Wilmott}\ }\textbf {\bibinfo {volume} {2022}},\ \bibinfo {pages} {100} (\bibinfo {year} {2022})}\BibitemShut {NoStop}%
\bibitem [{\citenamefont {Ledoit}\ and\ \citenamefont {P{\'e}ch{\'e}}(2011)}]{ledoit2011eigenvectors}%
  \BibitemOpen
  \bibfield  {author} {\bibinfo {author} {\bibfnamefont {O.}~\bibnamefont {Ledoit}}\ and\ \bibinfo {author} {\bibfnamefont {S.}~\bibnamefont {P{\'e}ch{\'e}}},\ }\bibfield  {title} {\bibinfo {title} {Eigenvectors of some large sample covariance matrix ensembles},\ }\href {https://doi.org/10.1007/s00440-010-0298-3} {\bibfield  {journal} {\bibinfo  {journal} {Probability Theory and Related Fields}\ }\textbf {\bibinfo {volume} {151}},\ \bibinfo {pages} {233} (\bibinfo {year} {2011})}\BibitemShut {NoStop}%
\bibitem [{\citenamefont {Alt}\ \emph {et~al.}(2020)\citenamefont {Alt}, \citenamefont {Erd\H{o}s},\ and\ \citenamefont {Kr{\"u}ger}}]{Alt2020energy}%
  \BibitemOpen
  \bibfield  {author} {\bibinfo {author} {\bibfnamefont {J.}~\bibnamefont {Alt}}, \bibinfo {author} {\bibfnamefont {L.}~\bibnamefont {Erd\H{o}s}},\ and\ \bibinfo {author} {\bibfnamefont {T.}~\bibnamefont {Kr{\"u}ger}},\ }\bibfield  {title} {\bibinfo {title} {The {Dyson} equation with linear self-energy: spectral bands, edges and cusps},\ }\href@noop {} {\bibfield  {journal} {\bibinfo  {journal} {Documenta Mathematica}\ ,\ \bibinfo {pages} {1421}} (\bibinfo {year} {2020})}\BibitemShut {NoStop}%
\bibitem [{\citenamefont {Cipolloni}\ \emph {et~al.}(2022)\citenamefont {Cipolloni}, \citenamefont {Erd\H{o}s},\ and\ \citenamefont {Schr{\"o}der}}]{Cipolloni2022Optimal}%
  \BibitemOpen
  \bibfield  {author} {\bibinfo {author} {\bibfnamefont {G.}~\bibnamefont {Cipolloni}}, \bibinfo {author} {\bibfnamefont {L.}~\bibnamefont {Erd\H{o}s}},\ and\ \bibinfo {author} {\bibfnamefont {D.}~\bibnamefont {Schr{\"o}der}},\ }\bibfield  {title} {\bibinfo {title} {{Optimal multi-resolvent local laws for Wigner matrices}},\ }\href {https://doi.org/10.1214/22-EJP838} {\bibfield  {journal} {\bibinfo  {journal} {Electronic Journal of Probabability}\ }\textbf {\bibinfo {volume} {27}},\ \bibinfo {pages} {1} (\bibinfo {year} {2022})}\BibitemShut {NoStop}%
\bibitem [{Note1()}]{Note1}%
  \BibitemOpen
  \bibinfo {note} {To the best of our knowledge, the two-resolvent local law \protect \eqref {eq:2Glaw} for sample covaraince matrices was first established in \cite {lin2024eigenvector}: their result is suboptimal and restricted to non-overlapping samples, and the error is suboptimal. However, with our dynamical proof techniques \cite {erdHos2024eigenstate}, the optimal result is attainable even in the correlated case \cite {erdHos2024cusp}.}\BibitemShut {Stop}%
\bibitem [{\citenamefont {Cipolloni}\ \emph {et~al.}(2024)\citenamefont {Cipolloni}, \citenamefont {Erd\H{o}s}, \citenamefont {Henheik},\ and\ \citenamefont {Schr{\"o}der}}]{Cipolloni2022overlap}%
  \BibitemOpen
  \bibfield  {author} {\bibinfo {author} {\bibfnamefont {G.}~\bibnamefont {Cipolloni}}, \bibinfo {author} {\bibfnamefont {L.}~\bibnamefont {Erd\H{o}s}}, \bibinfo {author} {\bibfnamefont {J.}~\bibnamefont {Henheik}},\ and\ \bibinfo {author} {\bibfnamefont {D.}~\bibnamefont {Schr{\"o}der}},\ }\bibfield  {title} {\bibinfo {title} {{Optimal lower bound on eigenvector overlaps for non-Hermitian random matrices}},\ }\bibfield  {journal} {\bibinfo  {journal} {Journal of Functional Analysis}\ }\href {https://doi.org/10.1016/j.jfa.2024.110495} {10.1016/j.jfa.2024.110495} (\bibinfo {year} {2024})\BibitemShut {NoStop}%
\bibitem [{\citenamefont {Mar{\v{c}}enko}\ and\ \citenamefont {Pastur}(1967)}]{marvcenko1967distribution}%
  \BibitemOpen
  \bibfield  {author} {\bibinfo {author} {\bibfnamefont {V.~A.}\ \bibnamefont {Mar{\v{c}}enko}}\ and\ \bibinfo {author} {\bibfnamefont {L.~A.}\ \bibnamefont {Pastur}},\ }\bibfield  {title} {\bibinfo {title} {Distribution of eigenvalues for some sets of random matrices},\ }\href {https://doi.org/10.1070/SM1967v001n04ABEH001994} {\bibfield  {journal} {\bibinfo  {journal} {Mathematics of the USSR-Sbornik}\ }\textbf {\bibinfo {volume} {1}},\ \bibinfo {pages} {457} (\bibinfo {year} {1967})}\BibitemShut {NoStop}%
\bibitem [{\citenamefont {Biroli}\ \emph {et~al.}(2007)\citenamefont {Biroli}, \citenamefont {Bouchaud},\ and\ \citenamefont {Potters}}]{biroli2007student}%
  \BibitemOpen
  \bibfield  {author} {\bibinfo {author} {\bibfnamefont {G.}~\bibnamefont {Biroli}}, \bibinfo {author} {\bibfnamefont {J.-P.}\ \bibnamefont {Bouchaud}},\ and\ \bibinfo {author} {\bibfnamefont {M.}~\bibnamefont {Potters}},\ }\bibfield  {title} {\bibinfo {title} {The {Student} ensemble of correlation matrices: eigenvalue spectrum and {Kullback-Leibler} entropy},\ }\href@noop {} {\bibfield  {journal} {\bibinfo  {journal} {arXiv preprint arXiv:0710.0802}\ } (\bibinfo {year} {2007})}\BibitemShut {NoStop}%
\bibitem [{\citenamefont {Espana}\ \emph {et~al.}(2024)\citenamefont {Espana}, \citenamefont {Coz},\ and\ \citenamefont {Smerlak}}]{espana2024kendall}%
  \BibitemOpen
  \bibfield  {author} {\bibinfo {author} {\bibfnamefont {T.}~\bibnamefont {Espana}}, \bibinfo {author} {\bibfnamefont {V.~L.}\ \bibnamefont {Coz}},\ and\ \bibinfo {author} {\bibfnamefont {M.}~\bibnamefont {Smerlak}},\ }\bibfield  {title} {\bibinfo {title} {Kendall correlation coefficients for portfolio optimization},\ }\href@noop {} {\bibfield  {journal} {\bibinfo  {journal} {arXiv preprint arXiv:2410.17366}\ } (\bibinfo {year} {2024})}\BibitemShut {NoStop}%
\bibitem [{\citenamefont {Lin}\ and\ \citenamefont {Pan}(2024)}]{lin2024eigenvector}%
  \BibitemOpen
  \bibfield  {author} {\bibinfo {author} {\bibfnamefont {Z.}~\bibnamefont {Lin}}\ and\ \bibinfo {author} {\bibfnamefont {G.}~\bibnamefont {Pan}},\ }\href@noop {} {\bibinfo {title} {Eigenvector overlaps in large sample covariance matrices and nonlinear shrinkage estimators}} (\bibinfo {year} {2024}),\ \Eprint {https://arxiv.org/abs/2404.18173} {2404.18173} \BibitemShut {NoStop}%
\bibitem [{\citenamefont {Erd{\H{o}}s}\ and\ \citenamefont {Riabov}(2024)}]{erdHos2024eigenstate}%
  \BibitemOpen
  \bibfield  {author} {\bibinfo {author} {\bibfnamefont {L.}~\bibnamefont {Erd{\H{o}}s}}\ and\ \bibinfo {author} {\bibfnamefont {V.}~\bibnamefont {Riabov}},\ }\bibfield  {title} {\bibinfo {title} {Eigenstate thermalization hypothesis for {Wigner-type} matrices},\ }\href {https://doi.org/10.1007/s00220-024-05143-y} {\bibfield  {journal} {\bibinfo  {journal} {Communications in Mathematical Physics}\ }\textbf {\bibinfo {volume} {405}},\ \bibinfo {pages} {282} (\bibinfo {year} {2024})}\BibitemShut {NoStop}%
\bibitem [{\citenamefont {Erd{\H{o}}s}\ \emph {et~al.}(2024)\citenamefont {Erd{\H{o}}s}, \citenamefont {Henheik},\ and\ \citenamefont {Riabov}}]{erdHos2024cusp}%
  \BibitemOpen
  \bibfield  {author} {\bibinfo {author} {\bibfnamefont {L.}~\bibnamefont {Erd{\H{o}}s}}, \bibinfo {author} {\bibfnamefont {J.}~\bibnamefont {Henheik}},\ and\ \bibinfo {author} {\bibfnamefont {V.}~\bibnamefont {Riabov}},\ }\href@noop {} {\bibinfo {title} {Cusp universality for correlated random matrices}} (\bibinfo {year} {2024}),\ \Eprint {https://arxiv.org/abs/2410.06813} {2410.06813} \BibitemShut {NoStop}%
\end{thebibliography}%

\section*{Supplementary Material}

	
	

\subsection*{Step 1. Computing the self-consistent resolvent}
In this subsection, we derive an explicit expression for $M(z)$. First, we compute the action of the self-energy operator. 

Consider an $L \times L$ (with $L$ given in \eqref{eq:L_def}) matrix $A$ with the following block structure,
\begin{equation}
    \begin{split}
        A &:= \begin{pmatrix}
    		\mathcal{A} & \star &\star\\
    		\star & \mathcal{D} &\star\\
    		\star &\star & \mathcal{E}
    	\end{pmatrix}, \quad \mathcal{A} := \begin{pmatrix}
    		\mathcal{A}_{11} & \mathcal{A}_{12}\\
    		\mathcal{A}_{21} & \mathcal{A}_{22}
    	\end{pmatrix},
    	\\ \mathcal{D} &:= \begin{pmatrix}
    		\mathcal{D}_{11} & \star\\
    		\star & \mathcal{D}_{22}
    	\end{pmatrix},
    	\quad \mathcal{E} := \begin{pmatrix}
    		\mathcal{E}_{11} & \mathcal{E}_{12}\\
    		\mathcal{E}_{21} & \mathcal{E}_{22}
    	\end{pmatrix},
    \end{split}
\end{equation}
where the blocks $\mathcal{A}_{jk}$ are $N\times N$, $\mathcal{D}_{11}$ is $T \times T$, $\mathcal{D}_{22}$ is $\wt T \times \wt T$, and $\mathcal{E}_{jk}$ are $T_B \times T_B$. All blocks denoted by $\star$ are irrelevant for the computation. Then, if follows form the definition of self-energy $S[\,\cdot\,]$ in \eqref{eq:self-energy} that
\begin{equation} \label{eq:S}
	S[A] = S_0[A] +  S_B[A],
\end{equation}
where the matrices $S_0[A]$ and $S_B[A]$ have the following block structures
\begin{equation} \label{eq:S0andS1}
    \begin{split}
        S_0[A] &= \begin{pmatrix}
		\mathcal{T}_0[\mathcal{D}] & \star  & \star\\
		\star & \mathcal{R}_0[\mathcal{A}] & \star\\
		\star & \star & 0
	\end{pmatrix},
	\\
	S_B[A] &= \begin{pmatrix}
		\mathcal{T}_B[\mathcal{E}] & \star  & \star\\
		\star & 0 & \star\\
		\star & \star & \mathcal{R}_B[\mathcal{A}]
	\end{pmatrix}.
    \end{split}
\end{equation}
Here block-trace operators $\mathcal{T}_{0/B}$ and $\mathcal{R}_{0/B}$ act by
\begin{equation}
    \begin{split}
        \mathcal{T}_0[\mathcal{D}] &= \begin{pmatrix}
		q\,\tau [ \mathcal{D}_{11}] & 0\\
		0 & \oq \,\tau [\mathcal{D}_{22}]
	\end{pmatrix} \otimes C,
	\\ 
	\mathcal{R}_0[\mathcal{A}] &= \begin{pmatrix}
		q\, \tau [ \mathcal{A}_{11} C ] I_T & 0\\
		0 & \oq  \,\tau [\mathcal{A}_{22}C ] I_{\wt{T}}
	\end{pmatrix} ,
    \end{split}
\end{equation}
\begin{equation} \label{eq:T1R1_def}
    \begin{split}
	\mathcal{T}_B[\mathcal{E}] &= \begin{pmatrix}
		q \, \tau [ \mathcal{E}_{11}]]& t{q \oq }\,\tau [ \mathcal{E}_{12}]\\
		\sqrt{q \oq }\, \tau [ \mathcal{E}_{21}] & \oq \,\tau [ \mathcal{E}_{22}]
	\end{pmatrix} \otimes C,
	\\ 
	\mathcal{R}_B[\mathcal{A}] &= \begin{pmatrix}
		q\, \tau [ \mathcal{A}_{11} C ] & \sqrt{q \oq }\,\tau [ \mathcal{A}_{12} C ]\\
		\sqrt{q \oq }\,\tau [ \mathcal{A}_{21} C ] & \oq \,\tau [ \mathcal{A}_{22}C ]
	\end{pmatrix} \otimes I_{T_B}.
    \end{split}
\end{equation}
Recall the convention $\tau [\cdot] := N^{-1} \Tr[\,\cdot\,]$.
The off-diagonal blocks in \eqref{eq:S0andS1} denoted by $\star$ arise from expectations of the form $\Expv[X (\cdot) X]$ and depend only on the $\star$ blocks of $A$. Indeed, it is easy to see that these blocks are irrelevant for the computation of $M(z)$, e.g., since the solution to the equation \eqref{eq:MDE} is unique, and these block vanish identically in the complex-Hermitian symmetry class.

By using the expressions \eqref{eq:S}--\eqref{eq:T1R1_def}, the equation \eqref{eq:MDE} can be reduced to two autonomous systems given by Eqs. \eqref{eq:Ms} in the main text. Indeed, it is straightforward to check that any solution to \eqref{eq:Ms} satisfying the positivity condition for the imaginary part induces a solution to \eqref{eq:MDE}. On the other hand, both systems \eqref{eq:Ms} (equivalent up to the change of parameter $q \leftrightarrow \oq $ ) admit a unique solution with positive imaginary part, since they correspond to the matrix Dyson equations associated with the random matrices of the form
\begin{equation}
	\begin{pmatrix}
		0 & X & B\\
		X^\top & 0 & 0\\
		B^\top & 0 & 0
	\end{pmatrix} \qquad \text{and} \qquad 
	\begin{pmatrix}
		0 & \wt{X} & B\\
		\wt{X}^\top & 0 & 0\\
		B^\top & 0 & 0
	\end{pmatrix}.
\end{equation}

\subsection*{Step 2. Computing the two-resolvent approximation}
We find $M^P$ using \eqref{eq:M12_def} in the block form
\begin{equation}
    \begin{split}
	M^P &:= \begin{pmatrix}
		\mathcal{F} & 0 & 0\\
		0 & \star & 0\\
		0 & 0 & \mathcal{J}
	\end{pmatrix}, \\
	\mathcal{F} &:= \begin{pmatrix}
		\mathcal{F}_{11} & \mathcal{F}_{12}\\
		\mathcal{F}_{21} & \mathcal{F}_{22}
	\end{pmatrix}, \quad 
	\mathcal{J} := \begin{pmatrix}
		\mathcal{J}_{11} & \mathcal{J}_{12}\\
		\mathcal{J}_{21} & \mathcal{J}_{22}
	\end{pmatrix},    
    \end{split}
\end{equation}
where the blocks $\mathcal{F}_{jk}$ are $N\times N$ and $\mathcal{J}_{jk}$ are $T_B\times T_B$. If fact, we only need to compute $\mathcal{F}_{12}$, since it is equal to the projection of $M^P$ onto $P$, defined in \eqref{eq:full_overlap}.  Using the explicit expression \eqref{eq:S} for the self-energy, we deduce the following systems of equations for $\mathcal{F}_{12}$ and $\mathcal{F}_{21}$,
\begin{equation} \label{eq:F12}
	\left\{ 
	\begin{array}{l}
		\smallskip
		\mathcal{F}_{12} - \sqrt{q \oq }\, \tau [ \mathcal{J}_{12} ]\, \mathcal{M} C \wt{\mathcal{M}} = \mathcal{M}\wt{\mathcal{M}} \\
		\smallskip
		\mathcal{J}_{12} - \sqrt{q \oq }\, \tau [ \mathcal{F}_{12}C] \m \om \, I_{T_B} = 0\\
	\end{array} 
	\right. ~,
\end{equation}
where we abbreviate $\M := \mathcal{M}(w)$, $\m := \m(w)$, and $\oM := \oM(\wt{w})$, $\om := \om(\wt{w})$.
We then immediately deduce Eq. \eqref{eq:F12_expr} in the main text. 

\end{document}